\documentclass[preprint]{aastex61}
\hypersetup{linkcolor=red,citecolor=blue,filecolor=cyan,urlcolor=magenta}

\shorttitle{A search for technosignatures in the Kepler field with the GBT}
\shortauthors{Margot et al. 2018}
\received{2018 February 3}
\revised{2018 March 3}
\accepted{2018 March 30}

\begin{document}

\title{A search for technosignatures from 14 planetary systems in the Kepler field with the Green Bank Telescope at 1.15--1.73 GHz}

\correspondingauthor{Jean-Luc Margot}
\email{jlm@astro.ucla.edu}

\author[0000-0001-9798-1797]{Jean-Luc Margot}
\affiliation{Department of Earth, Planetary, and Space Sciences, University of California, Los Angeles, CA 90095, USA}
\affiliation{Department of Physics and Astronomy, University of California, Los Angeles, CA 90095, USA}

\author{Adam H. Greenberg}
\affiliation{Department of Physics and Astronomy, University of California, Los Angeles, CA 90095, USA}

\author{Pavlo Pinchuk}
\affiliation{Department of Physics and Astronomy, University of California, Los Angeles, CA 90095, USA}

\author{Akshay Shinde}
\author{Yashaswi Alladi}
\affiliation{Department of Computer Science, University of California, Los Angeles, CA 90095, USA}

\author{Srinivas Prasad MN}
\affiliation{Department of Electrical Engineering, University of California, Los Angeles, CA 90095, USA} 

\author{M. Oliver Bowman}
\affiliation{Department of Earth, Planetary, and Space Sciences, University of California, Los Angeles, CA 90095, USA}

\author{Callum Fisher}
\affiliation{Department of Physics and Astronomy, University of California, Los Angeles, CA 90095, USA}

\author{Szilard Gyalay}
\affiliation{Department of Physics and Astronomy, University of California, Los Angeles, CA 90095, USA}

\author{Willow McKibbin}
\affiliation{Department of Physics and Astronomy, University of California, Los Angeles, CA 90095, USA}

\author{Brittany Miles}
\affiliation{Department of Physics and Astronomy, University of California, Los Angeles, CA 90095, USA}

\author{Donald Nguyen}
\affiliation{Department of Physics and Astronomy, University of California, Los Angeles, CA 90095, USA}

\author{Conor Power} %
\affiliation{Department of Electrical Engineering, University of California, Los Angeles, CA 90095, USA}

\author{Namrata Ramani}
\affiliation{Department of Materials Science and Engineering, University of California, Los Angeles, CA 90095, USA}

\author{Rashmi Raviprasad}
\affiliation{Department of Physics and Astronomy, University of California, Los Angeles, CA 90095, USA}

\author{Jesse Santana}
\affiliation{Department of Physics and Astronomy, University of California, Los Angeles, CA 90095, USA}

\author{Ryan S.\ Lynch}
\affiliation{Green Bank Observatory, PO Box 2, Green Bank, WV
  24494, USA}
\affiliation{Center for Gravitational Waves and Cosmology, Department
  of Physics and Astronomy, West Virginia University, White Hall, Box
  6315, Morgantown, WV 26506, USA}

\begin{abstract} %

  Analysis of Kepler mission data suggests that the Milky Way includes billions of Earth-like planets in the habitable zone of their host star.  Current technology enables the detection of technosignatures emitted from a large fraction of the Galaxy.  We describe a search for technosignatures that is sensitive to Arecibo-class transmitters located within $\sim$420 ly of Earth and transmitters that are 1000 times more effective than Arecibo within $\sim$13\,000 ly of Earth.
  Our observations focused on 14 planetary systems in the Kepler field and used the L-band receiver (1.15--1.73 GHz) of the 100~m diameter Green Bank Telescope.  Each source was observed for a total integration time of 5 minutes.  We obtained power spectra at a frequency resolution of 3~Hz and examined narrowband signals with Doppler drift rates between $\pm$9~Hz~s$^{-1}$.
We flagged any detection with a signal-to-noise ratio in excess of 10 as a candidate signal and identified approximately 850\,000 candidates.
  Most (99\%) of these candidate signals were automatically classified as human-generated radio-frequency interference (RFI).  A large fraction ($>$99\%) of the remaining candidate signals were also flagged as anthropogenic RFI because they have frequencies that overlap those used by global navigation satellite systems, satellite downlinks, or other interferers detected in heavily polluted regions of the spectrum.  All 19 remaining candidate signals were scrutinized and none were attributable to an extraterrestrial source.

\end{abstract}

\keywords{astrobiology --- extraterrestrial intelligence --- planets and satellites: general  --- (stars:) planetary systems --- techniques: spectroscopic} 
\section{Introduction}
Analysis of Kepler mission data suggests that the Milky Way includes billions of Earth-like planets in the habitable zone (HZ) of their host star \citep[e.g.,][]{boru16}.  The possibility that intelligent and communicative life forms developed on one or more of these worlds behooves us to conduct a search for extraterrestrial intelligence.
Here, we describe an L-band radio survey of 14 planetary systems selected from the Kepler mission field.
Our analysis methods are generally similar to those used by \citet{siem13}, but our observations sample a different slice of the search volume.  In addition, our analysis examines signals of lower signal-to-noise ratio (10 vs.\ 25) and larger range of Doppler drift rates ($\pm$9 Hz~s$^{-1}$ vs.\ $\pm$2 Hz~s$^{-1}$) than recent Breakthrough Listen results~\citep[][]{enri17}.

We define a ``technosignature'' as any measurable property or effect that provides scientific evidence of past or present technology, by analogy with ``biosignatures,'' which provide evidence of past or present life.  The detection of a technosignature such as an extraterrestrial signal with a time-frequency structure that cannot be produced by natural sources
would provide compelling evidence of the existence of another civilization.   A signal that is narrow ($<10$ Hz) in the frequency domain is a technosignature because natural sources do not emit such narrowband signals.  The narrowest reported natural emission spans about 550 Hz and corresponds to OH (1612 MHz) maser emission~\citep{cohe87}.  A monochromatic signal that shifts by $\pm$10 Hz as a function of time according to a complex sequence in a manner similar to that used in transmitting the 1974 Arecibo message \citep{naic75} is another technosignature.
This work focuses on detecting signals that are narrow in the frequency domain, and is sensitive to both of these examples.  Our data are also amenable to searching for signals that are narrow in the time domain (e.g., pulses).  

Our search is not predicated on the assumption of deliberate transmissions aimed at Earth.  Earthlings, for instance, use high-power ($\sim 10^6$ W) transmissions to study asteroids that may pose an impact hazard~\citep[e.g.,][]{naid16}.  These transmissions use monochromatic, binary phase-coded, or chirp signals, all of which would be recognized as technosignatures by alien civilizations.  In most such observations, less than a millionth of the energy is absorbed and scattered by the asteroid, and the remainder propagates beyond the asteroid at the speed of light.  Our search is agnostic about whether radio transmissions were intended for detection by a distant civilization (e.g., a beacon) or not (e.g., a radar or inter-planet telecommunication system).

Sections 2, 3, 4, and 5 describe the observations, analysis, discussion, and conclusions, respectively.

\section {Observations}
\label{sec-obs}
We selected 14 exoplanet host stars (Table \ref{tab-stars}) from the Kepler catalog.  A majority of these stars host small habitable zone planets with radii $R_p < 2 R_E$, where $R_E$ is Earth's radius \citep{kane16}.  Although such planets may be advantageous for the development of extraterrestrial life forms, advanced civilizations may be capable of thriving in a variety of environments, and we do not restrict our search to small habitable zone planets. Indeed, because planets and habitable zone planets are common among most stars in the Galaxy~\citep[e.g.,][]{peti13,bata14}, there is no compelling reason to search the Kepler field~\citep[e.g.,][]{siem13} as opposed to other fields.
There is, however, a possible increased probability of detecting technosignatures when observing planetary systems edge-on.
\begin{table}[h]
  \caption{Target host stars listed in order of observations.  Distances in parsecs are from the NASA Exoplanet Archive.  Habitable zone categories 1 and 2 refer to small ($R_p < 2 R_E$) planets in the conservative and optimistic habitable zones described by \citet{kane16}, respectively.  In multi-planet systems, only the lowest category is listed.}
  \label{tab-stars}
  \centering
\begin{tabular}{l c c}
  \hline  
    Host star  & Distance (ly) & HZ category \\  
 \hline   
Kepler-399 &   NA               & \\
Kepler-186 &   $561^{+42}_{-33}$ & Cat. 1              \\         
Kepler-452 &   NA               & Cat. 2              \\         
Kepler-141 &   NA               & \\
Kepler-283 &   NA               & Cat. 1               \\        
Kepler-22  &   620              &               \\  
Kepler-296 &   $737^{+91}_{-59}$ & Cat. 1                 \\      
Kepler-407 &   NA               & \\
Kepler-174 &   NA               & Cat. 2                 \\      
Kepler-62  &   1200             & Cat. 1       \\                
Kepler-439 &   $2260^{+215}_{-124}$  & \\
Kepler-438 &   $473^{+65}_{-75}$  & \\
Kepler-440 &   $851^{+52}_{-150}$  & Cat. 2             \\          
Kepler-442 &   $1115^{+62}_{-72}$  & Cat. 1               \\        
\hline
\end{tabular}

\end{table}

Our observing sequence was inspired by a solution to the traveling salesperson problem, which minimized the time spent repositioning the telescope.  Each target was observed twice in the following 4-scan sequence: target 1, target 2, target 1, target 2.  The integration time for each scan was $\tau = 150$~s, yielding a total integration time of 5 minutes per target.

We conducted our observations with the 100 m diameter Green Bank Telescope (GBT)~\citep{jewe04} on 2016 April 15, 16:00-18:00 Universal Time (UT).  We recorded both linear polarizations of the L-band receiver, which has a frequency range of 1.15--1.73 GHz.  Over this frequency range, the full-width half maximum (FWHM) beam width of the telescope is 11 arcmin--7.3 arcmin.  The aperture efficiency is $\sim$72\%,\footnote{https://science.nrao.edu/facilities/gbt/proposing/GBTpg.pdf} which provides an effective area of $\sim$5600 m$^2$ and telescope sensitivity of $\sim$2 K/Jy.\footnote{1 Jy = $10^{-26}$ ⁢W⁢m$^{-2}⁢$Hz$^{-1}$.}
At elevations above $\sim$20 degrees, the system temperature is $\sim$20~K and the system equivalent flux density (SEFD) is $\sim$10 Jy.
The L-band receiver is located at the Gregorian focus of the telescope, which was designed with an off-axis reflector to minimize stray radiation.

We used the GUPPI backend~\citep{guppi08}
in baseband recording mode and sampled 800~MHz of bandwidth from 1.1 to 1.9 GHz.
The signal was channelized into 256 channels of 3.125~MHz bandwidth each.  The raw voltages of the in-phase and quadrature channels were digitized with 8-bit quantization.  GUPPI's baseband recording mode enables reduction of the data storage requirements by a factor of four with minimal signal degradation with an optimal four-level (two-bit) sampler~\citep{koga98}\deleted{ implemented by field programmable gate arrays (FPGAs)}.  In this mode, the quantization thresholds are set to $-0.981599 \sigma$, 0, $+0.981599 \sigma$, where $\sigma$ is the root-mean-square (rms) of the voltage and the quantized levels are set to $\pm 1$ and $\pm 3.335875$.  The quantization efficiency, which is the ratio of signal power that is observed with the four-level sampler to the power that would be obtained with no quantization loss, is $\eta_Q = 0.8825$~\citep{koga98}. Eight computers handled the transfer of our data to 8 disk arrays at an aggregate rate of 800 MB/s or 2.88 TB/h.

Calibration procedures at the beginning of our observing window consisted of recording a monochromatic tone at 1501~MHz (center frequency + 1~MHz), performing a peak and focus procedure on a bright radio source near the Kepler field, and observing a bright pulsar near the Kepler field (PSR B2021+51) \citep{manc05}.

\section {Analysis}

\subsection{Validations}

We verified the validity of our data-processing pipeline by analyzing the monochromatic tone data and recovering the signal at the expected frequency.  We also folded the pulsar data at the known pulsar period and recovered the characteristic pulse profile.%

\subsection{Data selection}
We unpacked the data to 4-byte floating point values, computed Fourier transforms of the complex samples with the FFTW routine \citep{fftw05}, and calculated the signal power (Stokes I) at each frequency bin.  Signals with frequencies outside the range of the L-band receiver (1150--1730~MHz) were discarded.  We used the GBT's notch filter to mitigate interference from a nearby aircraft surveillance radar system.  Signals with frequencies within the 3 dB cutoff range of the notch filter (1200--1341.2~MHz) were also discarded.

\subsection{Bandpass correction}
Individual channels of the GUPPI instrument exhibited a mostly uniform bandpass response, with a few notable exceptions.  We fit a 16-degree Chebyshev polynomial to the median bandpass response of well-behaved channels.  We divided each power spectrum by this median response, which normalized signal levels across the entire bandpass.   

\subsection{Spectral analysis}
We used a Fourier transform length of $2^{20}$, corresponding to a time interval of $\Delta \tau = 0.336$ s and yielding a frequency resolution $\Delta f = 2.98$~Hz.  This choice of transform length and frequency resolution was dictated by our desire to examine drift rates of order 10 Hz~s$^{-1}$ (Section~\ref{sec-drift}). We stored about 450 consecutive power spectra, depending on the exact integration time of each scan, in frequency--time arrays of $2^{20}$ columns and 432--451 rows (hereafter, ``time-frequency diagrams'', sometimes known as ``spectrograms'' or ``spectral waterfalls'').  The average noise power was subtracted and the array values were scaled to the standard deviation of the noise power. 

\subsection{Drift rate analysis}
\label{sec-drift}
Because radio signals experience time-variable Doppler shifts due to the rotational and orbital motions of both emitters and observers, uncompensated signals smear in frequency space on a timescale $\Delta f/\dot{f}$, where $\Delta f$ is the frequency resolution and $\dot{f}$ is the Doppler drift rate.  The duration of a time series required to obtain a spectrum with resolution $\Delta f$ is $1/\Delta f$, such that the maximum drift rate that is observable without smear is $\dot{f}_{\rm max} =  \pm (\Delta f)^2$.  The maximum Doppler drift rates due to Earth's rotational and orbital motions are
$\sim$0.15 Hz~s$^{-1}$ and
$\sim$0.034 Hz~s$^{-1}$, respectively, at the maximum frequency of our observations and at Green Bank's latitude.
The Doppler drift rates due to the emitters are unknown.  We examined drift rates of up to $\dot{f}_{\rm max} = \pm (\Delta f)^2 = \pm 8.88$~Hz~s$^{-1}$, corresponding to accelerations of $\pm$1.6 m~s$^{-2}$, which admit a wide range of planetary radii, spin rates, orbital semi-major axes, and orbital periods.
We applied a de-smearing procedure to compensate for the accelerations of both emitter and observer.  Specifically, we implemented a computationally advantageous tree algorithm \citep{tayl74,siem13}, which enabled examination of 512 Doppler drift rates from 0 to 8.88 Hz~s$^{-1}$ in linearly spaced increments of
0.0173~Hz~s$^{-1}$.  This algorithm reads the frequency--time arrays, then shifts and sums
all powers corresponding to each of the 512 possible drift rate
values, effectively enabling integration of the signal power over the entire
scan duration without smear.  Application of the algorithm with
positive and negative Doppler drift rates resulted in two
frequency--drift rate arrays of $2^{20}$ columns and 512 rows for each
scan.

\subsection{Candidate signal detection}
We identified candidate signals with an iterative procedure.  We searched for the element with the highest signal-to-noise ratio (SNR) in the frequency--drift rate arrays.  The characteristics of this candidate signal (unique identifier, source name, scan number, scan start time, frequency at start of scan, drift rate, SNR, frequency resolution) were stored in a SQL database for subsequent analysis.  Because a candidate signal would often be detected redundantly at multiple drift rate values adjacent to that with the highest SNR, we decided to keep only the instance with the highest SNR value.  In order to do so, we blanked
the frequency--drift rate arrays in a region of frequency extent $\pm\dot{f}_{\rm max} \tau$ centered on the frequency of the highest SNR candidate signal.  We then repeated the procedure and searched for the element with the next highest SNR.  All candidate signals with SNR $>$ 10 were identified in this fashion and stored in the database.  We counted 858\,748 candidate signals, which amounts to $\sim$750\,000 candidates per hour of on-source integration time in the useful frequency range of the GBT L-band receiver.

\subsection{Rejection algorithms}
\label{sec-reject}
A signal from a distant source at rest or in uniform motion with respect to the observer exhibits no time variation in the value of the Doppler shift.  Signals from extraterrestrial sources, unless cleverly compensated for a specific location on Earth, experience a nonzero Doppler drift rate due in part to the rotational and orbital motions of Earth.  For these reasons, we categorized all signals with zero Doppler drift rate as likely terrestrial and eliminated them from further consideration.  About a quarter (231\,181) of the candidate signals were flagged on this basis, leaving 627\,567 candidates with nonzero Doppler drift rates.

To further distinguish between radio-frequency interference (RFI) and genuine extraterrestrial signals, we implemented two additional filters.  First, we flagged any signal that was not detected in both scans of the same source.  This filter can rule out many anthropogenic signals that temporarily enter the beam (e.g., satellite downlinks).  Second, we flagged any signal that appeared in more than one position on the sky.  This filter can rule out many anthropogenic signals that are detectable through the antenna sidelobes.  A logical AND was used to automatically flag candidate signals that remained for consideration after the rejection steps.   
Our rejection filters used the scan start times, frequencies, Doppler drift rates, and frequency resolutions stored in the SQL database to properly recognize signals from the same emitter observed at different times.
These filters successfully flagged 617\,410 of the remaining signals as likely anthropogenic, leaving 10\,157 signals for further investigation.

Overall, our rejection filters automatically eliminated 99\% of the initial detections as RFI.  

\subsection{Known interferers}
\label{sec-known}
\begin{figure}[h]
\centering
\noindent
\includegraphics[width=27pc]{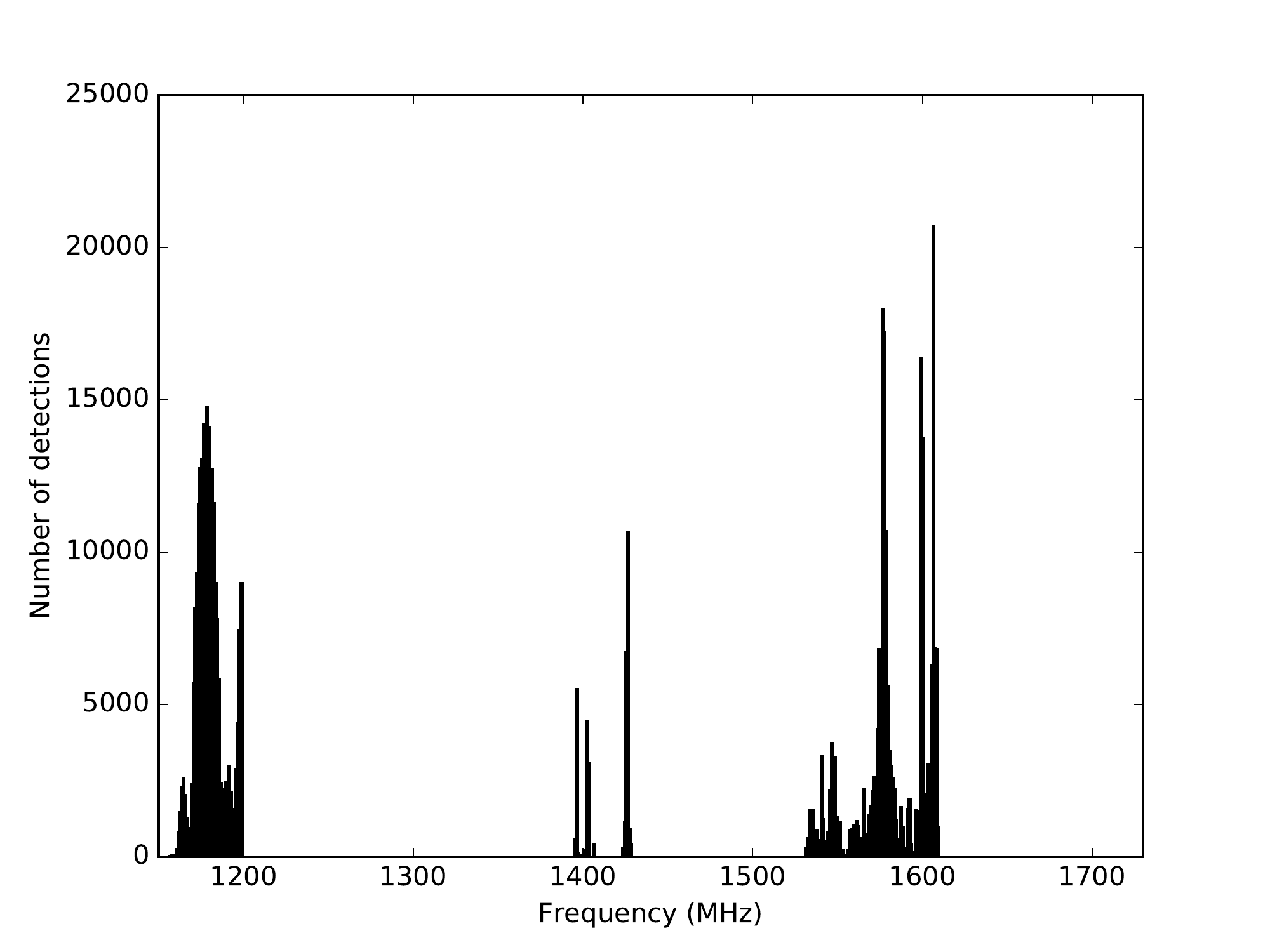}
\caption{Number of detections as a function of frequency, prior to application of our rejection algorithms (Section~\ref{sec-reject}).  Most of the high-density regions are due to the known interferers listed in Table~\ref{tab-rfi}.  The three clusters near 1400 MHz correspond to additional interferers described in Table~\ref{tab-morerfi}.}
\label{fig-rfi}  
\end{figure}

Several regions of the spectrum exhibit an unusually high density of detections (Figure~\ref{fig-rfi}).  Most of these high-signal-density regions can be attributed to known interferers.  We discarded all candidate signals in these regions, which we defined by the frequency extents shown in Table~\ref{tab-rfi}.  For signals generated by global navigation satellite systems, we used the signal modulation characteristics, typically binary phase-shift keying (BPSK), to delineate the frequency extent.  For satellite downlinks, we used the Federal Communications Commission (FCC) table of frequency allocations.
On average, the density of detections in the frequency regions ascribed to these satellites is $\sim$3\,500 detections per MHz of bandwidth.  In contrast, the density of signals in regions of the spectrum that exclude these interferers is $\sim$500 detections per MHz of bandwidth.

We flagged 9\,663 candidate signals out of 10\,157 as most likely due to these known interferers, leaving 494 signals for further consideration.

\begin{table}[h]
  \caption{Spectral regions exhibiting a high density of detections per unit frequency.  The number of detections are reported both prior to and after application of our rejection filters (Section~\ref{sec-reject}).  The density column shows the number of pre-filter detections per MHz.  Because some bands overlap, the totals are not the arithmetic sums of the table entries.}
\label{tab-rfi}
\centering
\begin{tabular}{ l l r r r l }
  \hline  
  Spectral region     & Width     & Pre-filter   & Post-filter  & Density        & Identification \\
  (MHz)               & (MHz)     & detections   & detections   & (\# per MHz)   &  \\   
 \hline                                                                                                      %
1554.96 -- 1595.88    & 40.92     & 107\,955     & 2657         & 2638           & GPS L1\\                  %
1155.99 -- 1196.91    & 40.92     & 206\,093     & 5289         & 5036           & GPS L5\\                  %
                                  
1592.9525 -- 1610.485 & 17.5325   & 80\,908      & 1110         & 4614           & GLONASS L1\\              %
1192.02   -- 1212.48  & 20.46     & 37\,792      & 518          & 1847           & GLONASS L3\\              %

1530 -- 1559          & 29        & 28\,814      & 89           & 994            &  Satellite downlinks\\    %
\hline                            
Total                 & 129.385   & 459\,543     & 9663         & 3552           & \\  %
\hline
\end{tabular}
\end{table}

\subsection{Additional interferers}
After we excised the known interferers listed in Table~\ref{tab-rfi}, we identified hundreds to thousands of detections with common time-frequency characteristics in narrow regions of the spectrum spanning a total of $\sim$9 MHz (Table~\ref{tab-morerfi}).  Because these classes of signals were each detected in at least six distinct pointing directions, they are almost certainly anthropogenic, and we flagged them as RFI.  We flagged 456 candidate signals out of 494 as most likely due to RFI, leaving 38 signals for further consideration.

\begin{figure}[h]
\centering
\noindent
\begin{tabular}{cc}
\includegraphics[width=21pc]{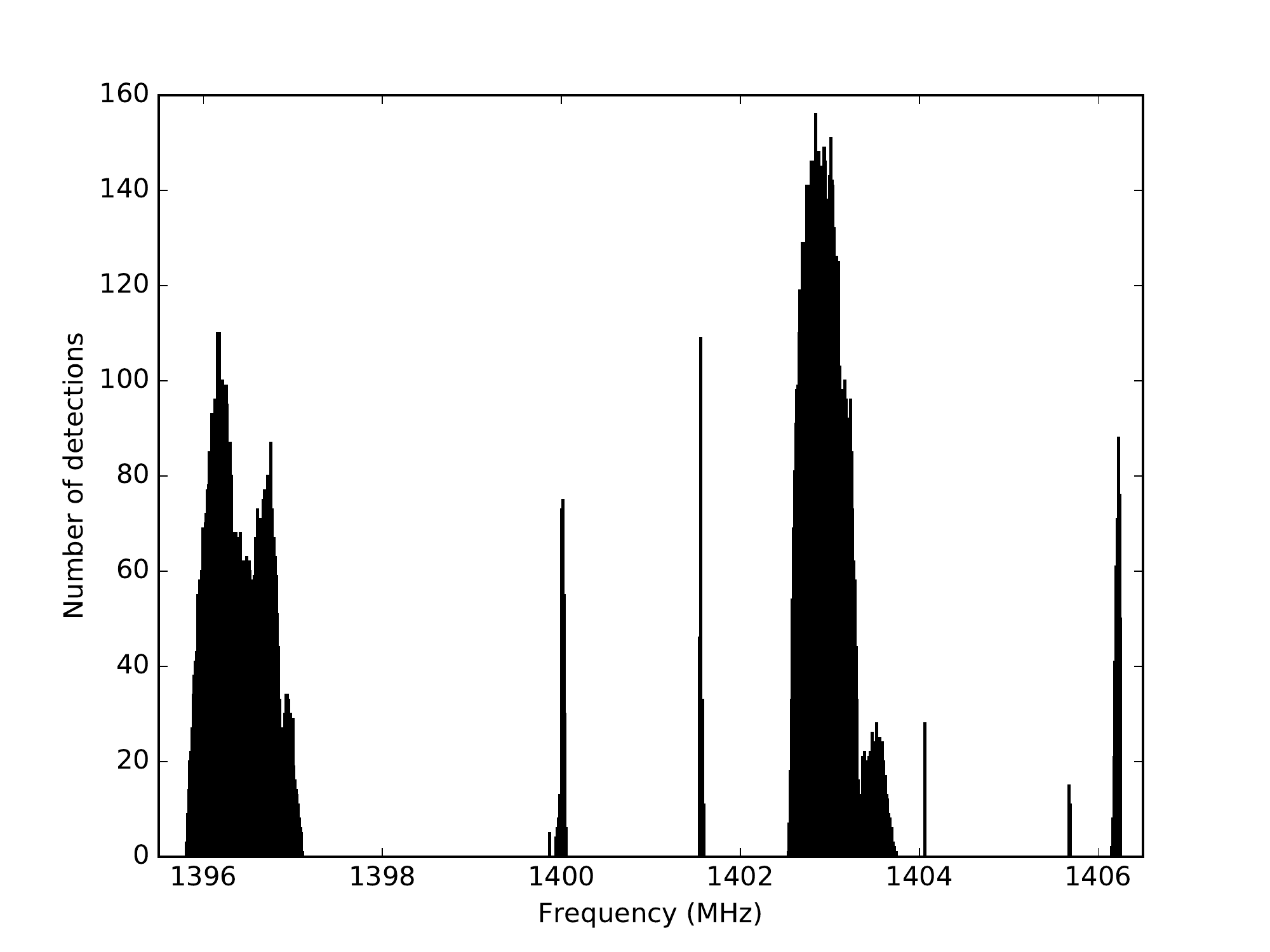}
&
\includegraphics[width=21pc]{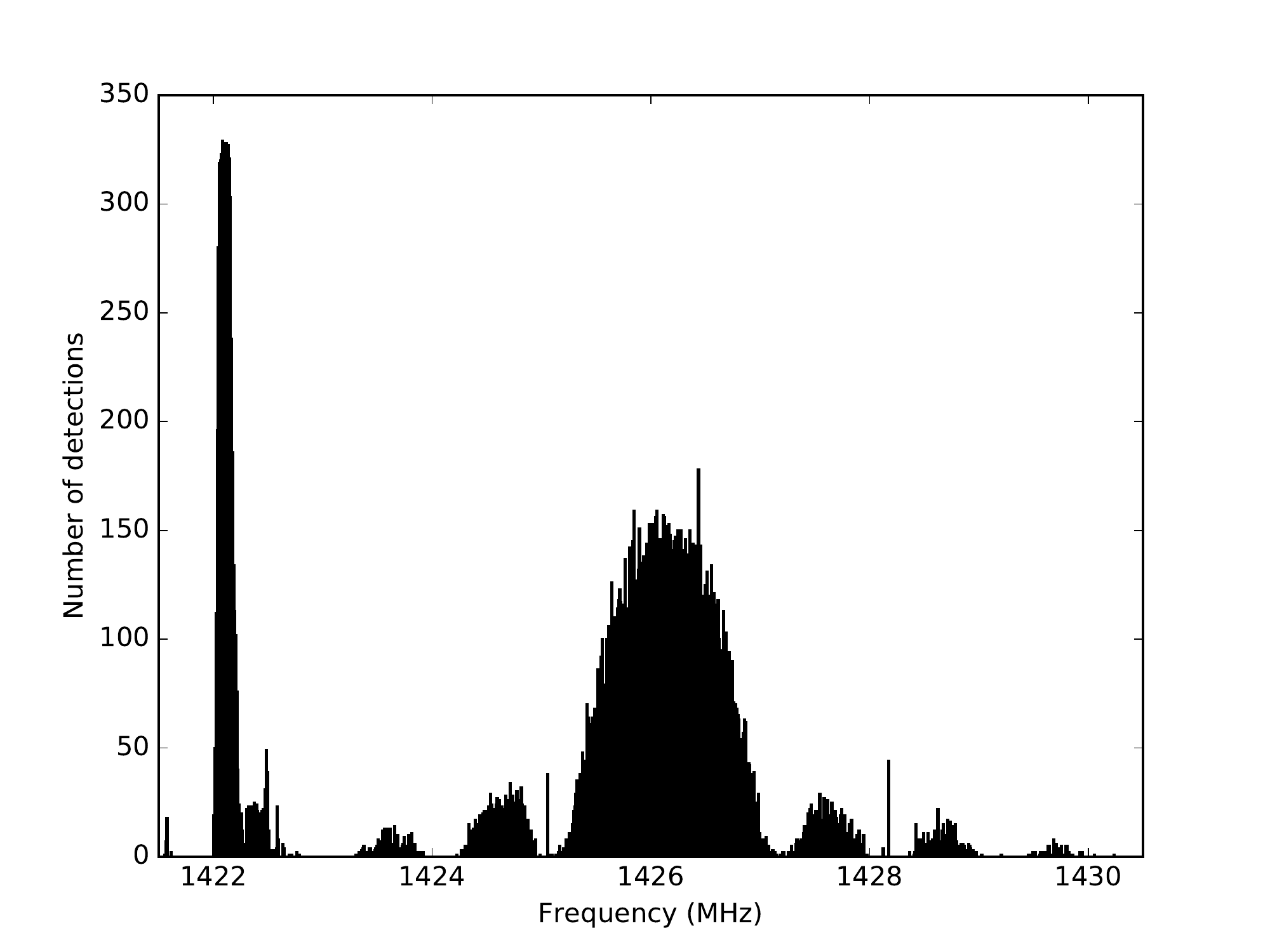}\\
\end{tabular}
\caption{Number of detections as a function of frequency, prior to the application of our rejection algorithms (Section~\ref{sec-reject}), in two 9 MHz-wide regions of the spectrum that partially overlap the 1400--1427~MHz radio astronomy protected band.  Signal characteristics are listed in Table~\ref{tab-morerfi}.  }
\label{fig-radio}  
\end{figure}

We detected strong interference in the radio astronomy protected band (1400--1427 MHz).  The interferers are visible in histograms of our detections in two 9-MHz-wide regions of the spectrum
(Figure~\ref{fig-radio}).  We briefly describe the characteristics of these interferers below.

The interferers near 1396 MHz, 1400 MHz, 1403 MHz, and 1406 MHz are similar.
They have a comb-like appearance in the frequency domain, with
spectral features spaced every kilohertz.  Each feature has a
bandwidth of about 140 Hz and resembles the modulation of a
double-sideband suppressed-carrier transmission.  In addition to these
features, strong carriers are observed at 1396.18~MHz, 1396.75 MHz,
and 1402.94 MHz.
Most of these interferers are in the radio astronomy protected band.

The region of the spectrum near 1401.5 MHz is characterized by a number of narrow lines 
that seem to cluster near 10 discrete
  frequency regions.  They have small (mostly $<$0.2 Hz s$^{-1}$)  
positive or negative Doppler drift  rates.
Some of these lines exhibit somewhat erratic behavior as a
function of time, perhaps indicating an unstable oscillator.  The
small Doppler drift rates suggest a terrestrial source.

The interferer near 1422 MHz generates a broad ($\sim$270 kHz) region
of increased noise power without distinct lines, making the
identification of the Doppler drift rate difficult.

The region between 1423 MHz and 1429 MHz exhibits some interferers
that are approximately 70~Hz wide and reminiscent of those described
by \citet[][their figure 5]{siem13}.  Others are time-variable signals
(12--15 s periodicity) that are approximately 70 Hz wide.  These
characteristics are similar to those of some Air Route Surveillance
Radars (ARSR) that track aircraft in all azimuthal directions at 5
rotations per minute, except that these systems are designed to
operate between 1250 MHz and 1350 MHz.  If due to ARSR, it is unclear
whether the interference is caused at the source or generated by
intermodulation products in the GBT receiver system.  The sinc-type
appearance of the histogram (Figure~\ref{fig-radio}) is reminiscent of
the power spectrum of a BPSK waveform with 1 MHz bandwidth and
compressed pulse width of 1 $\mu$s.  Such waveforms are used to
provide radar imaging with 150~m resolution~\citep[e.g.,][]{marg99a}.
Some ARSR systems use non-linear frequency modulation (NLFM) to
provide a nominal range resolution of 116~m.  Some NLFM schemes
exhibit a sinc-like power spectrum.  In addition to the periodic
interferers, strong carriers with variable drift rates are observed
near 1425.05 MHz and 1428.18~MHz.

\begin{table}[h]
\caption{Characteristics of likely anthropogenic interferers.  The number of detections are reported both prior to and after the application of our rejection filters (Section~\ref{sec-reject}).  Doppler drift rate and SNR statistics are computed on the pre-filter detections. Strong interferers are detected in the radio astronomy protected band (1400--1427 MHz).}
\label{tab-morerfi}
\centering
\begin{tabular}{ l l r r r c }
  \hline  
  Spectral region       & Width   & Pre-filter   & Post-filter  & Median drift  & SNR\\  
  (MHz)               & (MHz)     & detections   & detections   & rate (Hz~s$^{-1}$)   & (min/median/max)\\
 \hline                           
1395.810 -- 1397.097   & 1.287     & 6210         &  46         & 0.451 & 10.0/15.9/1087.2\\ %

1399.872 -- 1400.054   & 0.182     & 287          &   0          & 0.434 & 10.0/12.5/26.1\\ %
1401.547 -- 1401.599   & 0.052     & 220          &   2          & -0.017 & 10.0/19.0/1423.6\\ %
1402.536 -- 1403.750   & 1.214     & 7536         &  76          & 0.434 & 10.0/19.5/451.8\\ %
1406.145 -- 1406.250   & 0.105     & 418          &   4          & 0.434 & 10.0/12.6/25.6\\ %

1422.000 -- 1422.270   & 0.270     & 5150         &   2          &  N/A & 10.0/17.4/84.1\\ %

1423.308 -- 1428.971   & 5.663     & 20\,134      & 326          & 0.121 & 10.0/21.9/7642.2\\ %
\hline                   
Total                  & 8.773     & 39\,955      & 456          &          & \\
\hline
\end{tabular}
\end{table}

\begin{figure}[h]
\centering
\noindent
\includegraphics[width=6.5in]{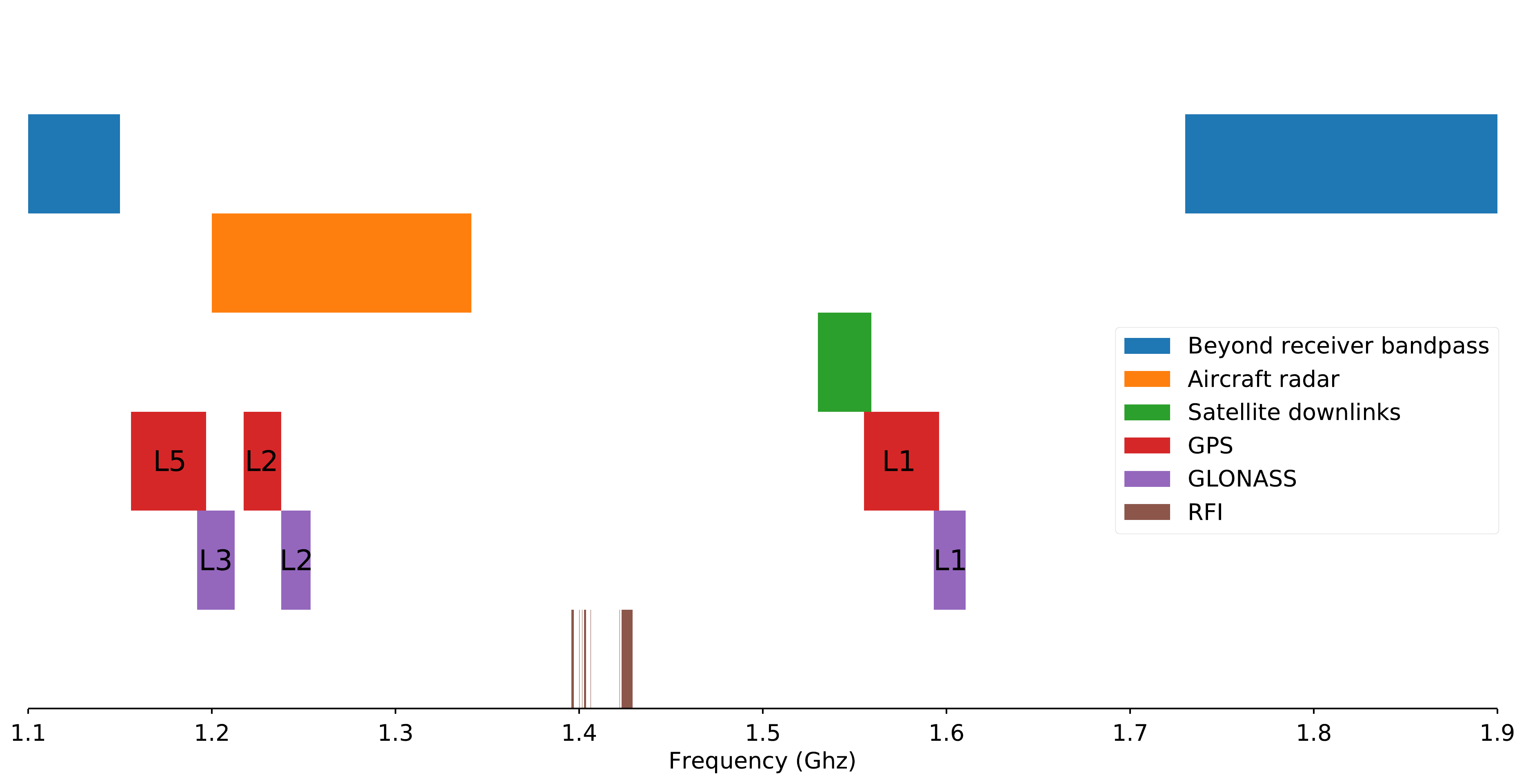} 
\caption{Color-coded summary of frequency regions that were excluded from our analysis.}
\label{fig-bands}  
\end{figure}

Figure~\ref{fig-bands} shows a graphical summary of the frequency regions that were excluded from our analysis.

\subsection{Evaluation of remaining candidates}
The 38 remaining signals represent 19 pairs of scans.  The characteristics of the first scans are shown in Table~\ref{tab-seti}.  We generated frequency-time diagrams for all 19 pairs.
We then searched our database for signals at similar frequencies and compared their time-frequency diagrams to those of our top candidates.  This process revealed that all 19 candidates are detected in more than one direction on the sky, ruling out the possibility of an extraterrestrial signal.  Examination of the time-frequency diagrams revealed groups of signals that can be attributed to the same source of RFI.  We provide a brief description of the signals below, along with a few examples of time-frequency diagrams.  To our knowledge, these interferers have not been reported in the literature, although they may of course have been detected in other searches.  

\begin{table}[h]
  \caption{Characteristics of top candidates listed in increasing order of frequency.  Epoch, frequency, and drift rate refer to the beginning of the first scan with units of modified Julian date (MJD), Hz, and Hz~s$^{-1}$, respectively.  SNR refers to the integrated power over the scan duration after correcting for the corresponding Doppler drift rate.}
\label{tab-seti}
\centering
\begin{tabular}{ r l c c r r}
  \hline
  ID      &       Source          &       Epoch             &       Frequency               &      Drift rate           & SNR  \\
          &                       &       (MJD)             &       (Hz)                    &      (Hz~s$^{-1}$)               &        \\  
  \hline  
128112	&	Kepler-283      &	57493.70692	&	1151551501.442434	&       -0.2429	&	21.6  \\ %
141584	&	Kepler-442      &	57493.74497	&	1375489655.604034	&	-8.8124	&	16.7  \\ %
2914	&	Kepler-141      &	57493.70009	&	1414058674.868273	&	 0.0867	&      544.0  \\ %
93976	&	Kepler-399      &	57493.68869	&	1444533390.553847	&	 0.0173	&	14.2  \\ %
36563	&	Kepler-186      &	57493.69100	&	1453857276.541974	&	 0.0173	&	16.0  \\
52175	&	Kepler-22       &	57493.70934	&	1453895304.341606	&	 0.0173	&	19.1  \\
81528	&	Kepler-296      &	57493.71608	&	1457414449.371766	&	 0.3123	&	18.9  \\
108574	&	Kepler-407      &	57493.71818	&	1457414503.015998	&	 0.3469	&	10.7  \\
81536	&	Kepler-296      &	57493.71608	&	1457442281.787187	&	 0.2949	&	14.5  \\
108572	&	Kepler-407      &	57493.71818	&	1457453120.902177	&	 0.3643	&	12.2  \\
81533	&	Kepler-296      &	57493.71608	&	1457490984.788880	&	 0.3123	&	16.7  \\
108848	&	Kepler-407      &	57493.71818	&	1461771970.293017	&	-0.0173	&	17.6  \\    %
11819	&	Kepler-141      &	57493.70009	&	1467993506.067759	&	-0.0173	&	13.3  \\ %
123438	&	Kepler-438      &	57493.73606	&	1472254848.842477	&	 0.0173	&	16.4  \\    %
72937	&	Kepler-283      &	57493.70692	&	1472262660.038624	&	-0.0347	&	13.8  \\
24383	&	Kepler-174      &	57493.72476	&	1472262722.623560	&	-0.0347	&	75.4  \\
153968	&	Kepler-440      &	57493.74274	&	1485092589.943495	&	-0.0173	&	10.4  \\ %
956243	&	Kepler-442      &	57493.74497	&	1501557959.611854	&	 0.1041	&	10.2  \\ %
1129207	&	Kepler-439      &	57493.73383	&	1623514653.815893	&	-0.1735	&	22.0  \\ %
\hline
\end{tabular}
\end{table}

\begin{figure}[p]
\centering
\noindent
\includegraphics[width=28pc]{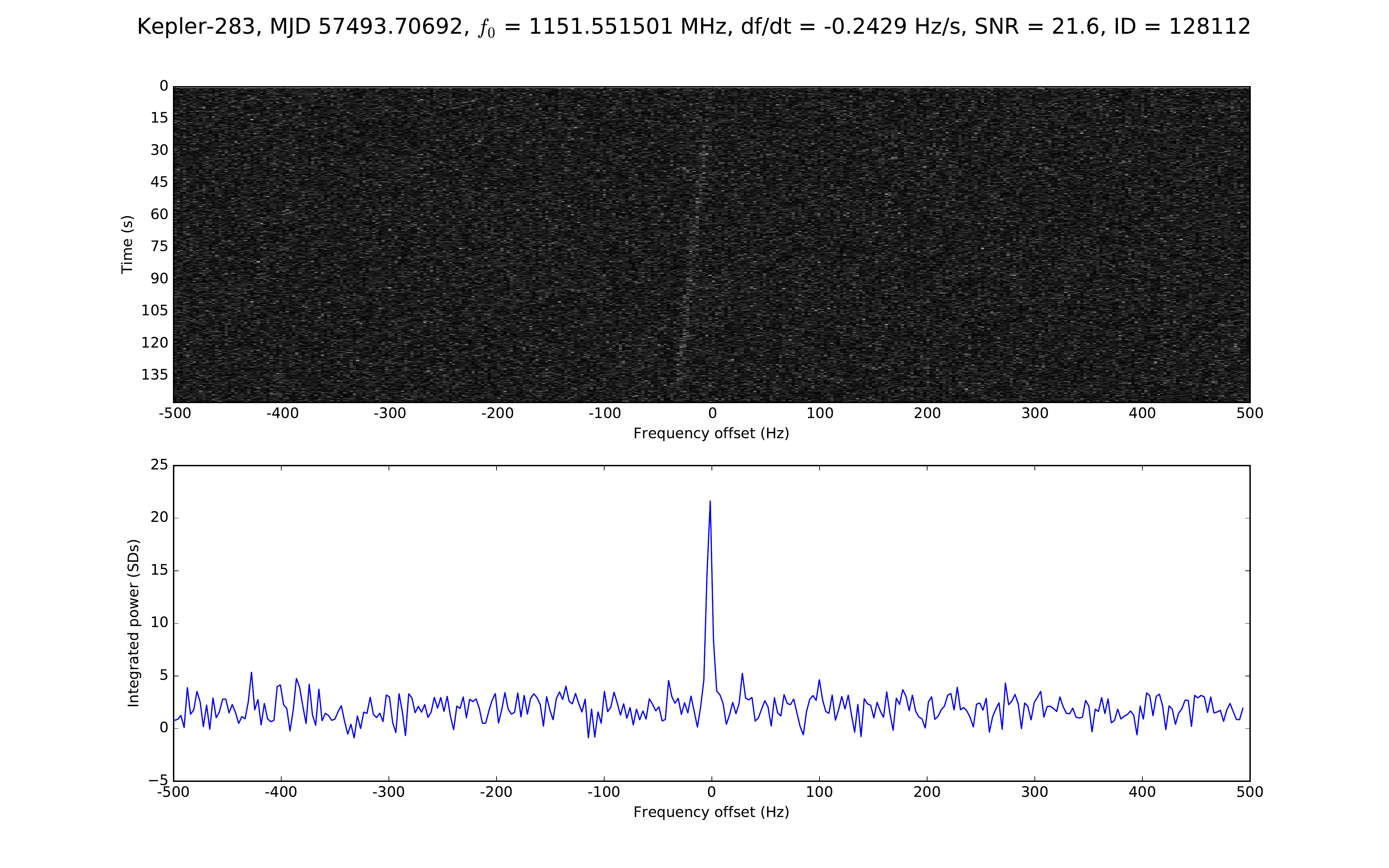} 
\includegraphics[width=28pc]{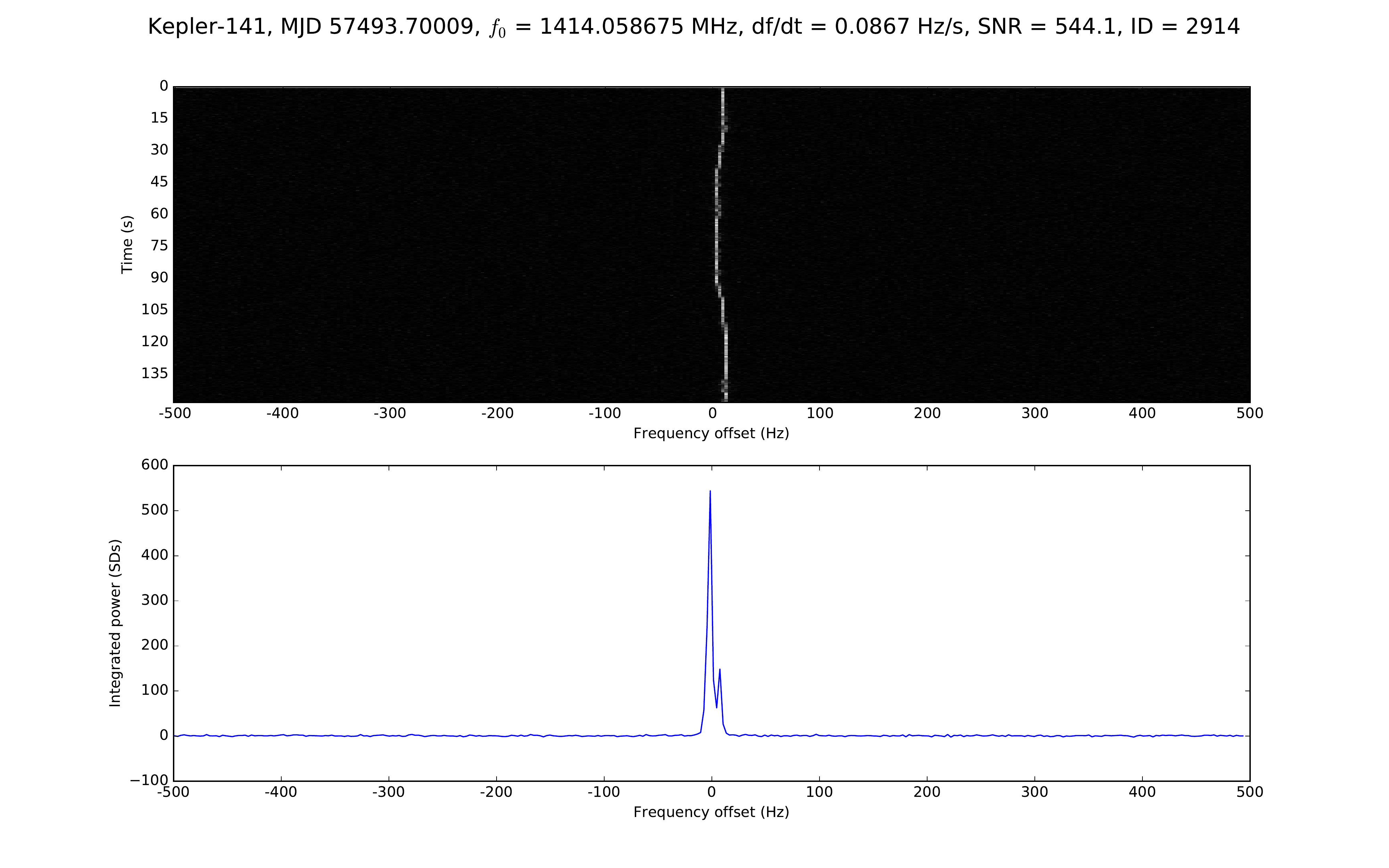} 
\includegraphics[width=28pc]{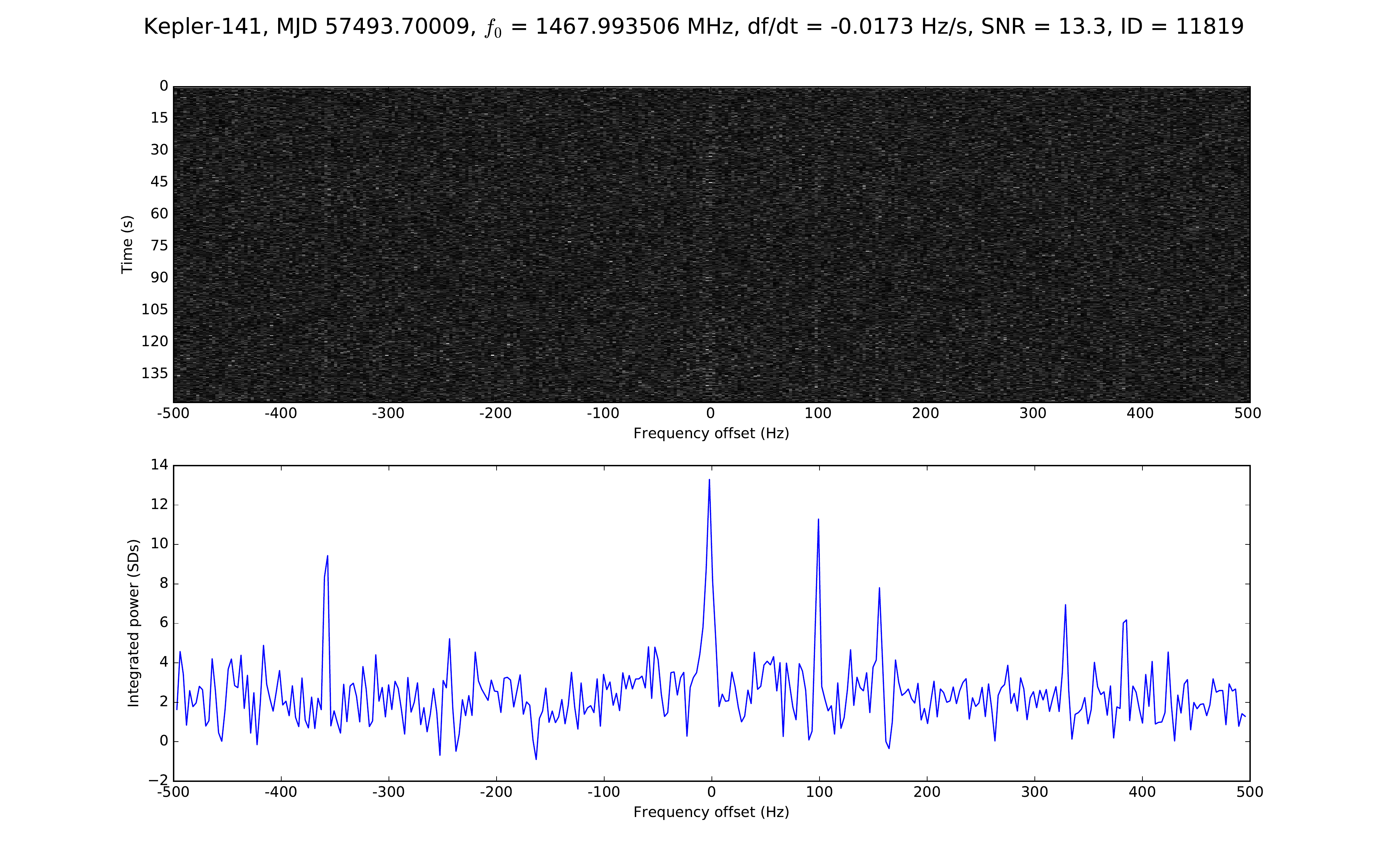} 
\caption{Time-frequency characteristics of three candidate signals. In each panel: (Top) Time-frequency diagrams showing consecutive power spectra; (Bottom) Integrated (i.e., shifted and summed) power spectra.}
\label{fig-cands}  
\end{figure}

The candidate signal near 1151 MHz (Figure~\ref{fig-cands}) is a monochromatic signal with a substantial Doppler drift rate that is observed in at least eight distinct directions on the sky.

The candidate signal near 1375 MHz has a complex time-frequency structure that is observed in at least one other direction on the sky.

The candidate signal near 1414 MHz (Figure~\ref{fig-cands}) exhibits both high SNR and somewhat erratic frequency behavior.
The rejection filter logic likely failed because the Doppler behavior of this signal is  erratic.

The candidate signal near 1444 MHz is a low SNR monochromatic signal that is observed in at least four other directions on the sky.

Both candidates near 1453 MHz exhibit a set of intermittent narrow lines that are observed in multiple directions on the sky.

All five candidates near 1457 MHz are due to anthropogenic RFI and share the same characteristics, i.e., a monochromatic signal superimposed on a broadband, elevated noise level.

The candidate signal near 1461 MHz is a single line that is observed in at least two directions on the sky.  Moreover, signals with similar lines that are offset in frequency by almost exactly 20 kHz and 30 kHz are observed in other directions on the sky.

The candidate signal near 1467 MHz (Figure~\ref{fig-cands}) exhibits a set of monochromatic lines.  The strongest one exceeded our SNR threshold.  It is observed in at least four directions on the sky.

The candidate signal near 1472.254 MHz is detected in at least three directions on the sky.

The candidates near 1472.262 MHz correspond to the same signal that is observed in at least two distinct directions.

The candidates near 1485 MHz and 1501 MHz are monochromatic signals that are detected in at least two and six directions on the sky, respectively.

The candidate signal near 1623 MHz has a complex time-frequency structure that is observed in several directions on the sky.  An approximately linear ramp with slope -150 Hz s$^{-1}$ appears then disappears with on and off durations of 5 to 10 s.

\section{Discussion}
\subsection{Search volume}
Although no extraterrestrial signals were identified in our analysis to date, we emphasize that our study encompassed only a small fraction of the search volume.  The fraction of the sky that was covered in our search is 14 times the solid angle of the GBT beam.
At the center frequency of our search, each beam has a solid angle of 0.015 deg$^2$.
Considering all 14 sources, we covered 0.21 deg$^2$ or about 5 ppm of the entire sky.  Our observations lasted a total of 5 minutes on each source, which is about
10 ppm of a terrestrial year.  After the elimination of polluted bands, our useful bandwidth spanned almost exactly 300~MHz, which is a small fraction of the electromagnetic spectrum available for transmission.

We computed the Drake figure of merit~\citep{drak84,enri17} that
corresponds to our search parameters:
\begin{equation}
    {\rm DFM} = \frac{\Delta f_{\rm tot} \Omega}{S_{\rm det}^{3/2}},
\end{equation}
where $\Delta f_{\rm tot}$ = 300 MHz is the total bandwidth examined,
$\Omega$ = 0.21 deg$^2$ is the total area of the sky covered, and
$S_{\rm det}$ = 10 Jy is the minimum flux density required for a
detection (Section~\ref{sec-sens}).  In these units, we find DFM
$\simeq 2\times 10^{6}$.  Referring to the values provided in
\citet{enri17}, our search is about 150 times larger than that of
\citet{horo93} and 10 times larger than that of \citet{gray17}.  It
amounts to about 2\% and 12\% of the recent large surveys by
\citet{enri17} and \citet{harp16}, respectively.

\subsection{Existence limits}
Attempting to place existence limits on the basis of SETI observations is a difficult exercise.  Certainly, we can place limits only on the kinds of signals that we are looking for, not actual limits on the presence of civilizations.  For instance, \citet{enri17} attempted to place a limit on the number of 100\%-duty cycle transmitters (e.g., a radio beacon) and suggested that fewer than 0.1\% of the stellar systems within 50 pc possess such transmitters.  However, beacons operating at frequencies lower than 1.1 GHz, larger than 1.9 GHz, or in the 1.2--1.34 GHz range would be undetected in their (and our) search, which makes general claims about the number of beacons unreliable.  In this spirit, we describe the types of signals that are detectable with our search, but we do not attempt to make inferences about the prevalence of radio beacons in the Galaxy.

\subsection{Sensitivity}
\label{sec-sens}
For the detection of narrowband signals above a floor with noise fluctuations, the SNR can be expressed as
$P_r/\Delta P_{\rm noise}$,
where $P_r$ is the received power and $\Delta P_{\rm noise}$ is the standard deviation of the receiver noise given by
\begin{equation}
  \Delta P_{\rm noise} = \frac{k_B T_{\rm sys} \Delta f}{\sqrt{\Delta f \tau}},
\end{equation}
  with $k_B$ Boltzmann's constant, $T_{\rm sys}$ the system temperature, $\Delta f$ the frequency resolution, and $\tau$ the integration time~\citep{ostr93,naid16}.  
  The power received by a transmitter of power $P_t$ and antenna gain $G_t$ located at distance $r$ is
  \begin{equation}
    P_r = \frac{P_t G_t}{4 \pi r^2} \frac{A_e}{2},
  \end{equation}
  where $A_e$ is the effective area of the receiving station and the factor of 1/2 accounts for reception by a single-polarization feed.  If powers received in both polarizations are added incoherently as in this work, ${\rm SNR} = \sqrt{n_{\rm pol}} \times P_r/\Delta P_{\rm noise}$, with $n_{\rm pol} = 2$.  The SNR is proportional to the product of factors that relate to the transmitter-receiver distance, transmitter performance, receiver performance, quantization efficiency, and data-taking and data-analysis choices, as shown in this expression:
  \begin{equation}
{\rm  SNR} = \left( \frac{1}{4\pi r^2} \right ) \left(PtGt\right) \left(\frac{A_e}{2 k_B T_{\rm sys}} \right) \left( \eta_Q \right) \left( {\frac{n_{\rm pol}\,\tau}{\Delta f}} \right)^{1/2}.
  \end{equation}
The second factor is known as the effective isotropic radiated power (EIRP) and the third factor is the inverse of the system-equivalent flux density (SEFD), so we can rewrite the SNR as
  \begin{equation}
{\rm  SNR} = 9\,  \left( \frac{100\ {\rm ly}}{r} \right )^2 \left(\frac{{\rm EIRP}}{10^{13}\ {\rm W}}\right) \left(\frac{10\ {\rm Jy}}{{\rm SEFD}} \right) \left( \frac{\eta_Q}{1} \right) \left( {\frac{n_{\rm pol}}{1}} \right)^{1/2} \left( {\frac{\tau}{1\ {\rm s}}} \right)^{1/2} \left( {\frac{1\ {\rm Hz}}{\Delta f}} \right)^{1/2}.    
\label{eq-snr}
  \end{equation} %
  The SNR is maximized when the frequency resolution of the data matches the bandwidth of the signal.

  We used the Arecibo planetary radar as a prototype transmitter with $P_t$ = 10$^6$ W, $G_t$ = 73.4 dB, and EIRP = 2.2 $\times 10^{13}$ W.
  The nominal receiver is the GBT with $A_e$ = 5600 m$^2$, $T_{\rm sys}$ = 20~K, and SEFD = 10 Jy.  The quantization efficiency of the four-level sampler is $\eta_Q = 0.8825$.  Our data-taking and data-analysis choices correspond to $n_{\rm pol} = 2$, $\tau$ = 150~s, and $\Delta f$ = 3~Hz.  With these values, we find $\Delta P_{\rm noise} = 3.9\times 10^{-23}$ W and $P_r = 3.1 \times 10^{-22}$ W for a transmitter located at a distance of 420~ly (128 pc) from Earth.  Such transmissions would be at the limit of detection with our minimum SNR threshold of 10.\footnote{An SNR threshold of 10 with our search parameters and $\eta_Q =1$ corresponds to detection of flux densities of 10 Jy.}

\begin{figure}[h]
\centering
\noindent
\includegraphics[width=30pc]{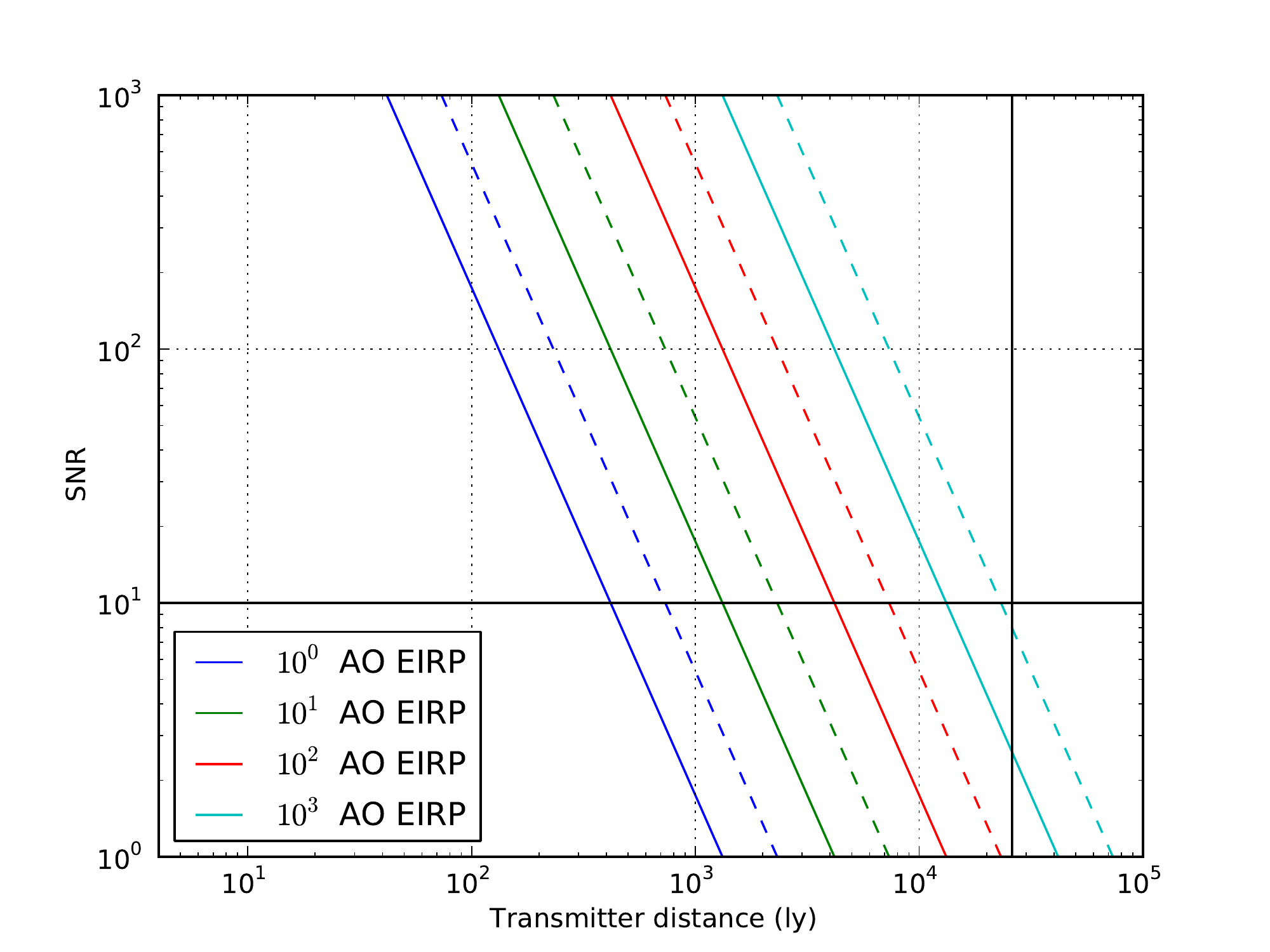}
\caption{SNR of detections as a function of transmitter distance from Earth, assuming search parameters identical to those used in this work ($n_{\rm pol} = 2$, $\tau$ = 150~s, and $\Delta f$ = 3~Hz).  Colored lines represent transmitter EIRPs that are equivalent to 1--1000 times that of the Arecibo Observatory (AO) planetary radar.  Solid and dashed colored lines represent reception with the GBT (SEFD = 10 Jy) and AO (SEFD = 3.2 Jy), respectively.  The black horizontal line represents the threshold for detection used in this work.  The black vertical line represents the distance to the center of the Galaxy.  }
\label{fig-snr}  
\end{figure}

The target stars with known distances in our sample are located at distances that exceed 420 ly, such that a more powerful transmitter, a more sensitive receiver, longer integration times, or narrower frequency resolutions would be needed for detection.  Transmitters with considerably larger EIRPs than that of Arecibo may be available to other civilizations.  A transmitter with 1000 times Arecibo's EIRP would be detectable in our search from distances of up to 13\,250~ly.  Current technology enables the detection of technosignatures emitted from a large fraction of the Galaxy (Figure~\ref{fig-snr}).  In such a vast search volume, there are billions of
 targets accessible to a search for technosignatures.  In contrast, the search for biosignatures will be limited in the foreseeable future to a few targets in the Solar System and to a few hundred planetary systems around nearby stars.

One can use Figure~\ref{fig-snr} to evaluate the detectability of the Arecibo planetary radar by other civilizations.  From an SNR standpoint, an increase by a factor of 10 in EIRP is equivalent to an increase in effective area by a factor of 10.  In other words, the colored lines in Figure~\ref{fig-snr} also represent the detectability of Arecibo by remote antennas with effective areas that are 1-1000 times larger than those of the GBT or Arecibo, assuming similar system temperatures and search parameters ($n_{\rm pol} = 2$, $\tau$ = 150~s, and $\Delta f$ = 3~Hz).  Equation~(\ref{eq-snr}) can be used to evaluate detectability with other transmitter, receiver, or search parameters.  Typical planetary radar transmissions have a duty cycle of approximately 50\%, and the maximum tracking duration for celestial sources at Arecibo is $\sim$2.5 h.  For observations of main belt asteroids and more distant bodies at opposition, the pointing direction over the entire tracking duration changes by less than 2 arcmin, i.e., less than the width of the beam for S-band (2.38 GHz) planetary radar transmissions.

\section {Conclusions}
\label{sec-conclusions}

We described the results of a search for narrowband signals from extraterrestrial sources using two hours of GBT telescope time in 2016.  We identified 858\,748 candidate signals.
Our rejection filters automatically eliminated 99\% of the candidates, leaving 10\,157 candidate signals for further inspection.  Almost all of the remaining signals were ruled out because they were attributable to anthropogenic RFI, leaving 19 pairs of candidate signals.  All of these candidates were observed in more than one direction on the sky, thereby ruling them out as extraterrestrial signals. 

Our observations were designed, obtained, and analyzed by students enrolled in a UCLA course titled ``Search for Extraterrestrial Intelligence: Theory and Applications.''
The search for technosignatures provides a superb educational opportunity for students in astrophysics, computer science, engineering, mathematics, planetary science, and statistics.  In this work, six graduate students and nine undergraduate students at UCLA learned valuable skills related to radio astronomy, telecommunications, programming, signal processing, and statistical analysis.  A course narrative is available at \href{http://seti.ucla.edu}{http://seti.ucla.edu}.

\section*{Acknowledgments}
We thank Janet Marott, Larry Lesyna, and David Saltzberg for the
financial support that made the 2016 observations and analysis
possible.  We thank Smadar Gilboa, Marek Grze\'skowiak, and Max
Kopelevich for providing an excellent computing environment in the
Orville L. Chapman Science Learning Center at UCLA.  We are grateful
to Wolfgang Baudler, Frank Ghigo, Ron Maddalena, Toney Minter, and
Karen O'Neil for assistance with the GBT observations, and to Phil
Perillat for RFI identification.  We are grateful to the designers and
funders of GUPPI for making the system available to us, with special
thanks to Paul Demorest and John Ford.  We are grateful to the
reviewer for useful comments. This research has made use of the NASA
Exoplanet Archive, which is operated by the California Institute of
Technology, under contract with the National Aeronautics and Space
Administration under the Exoplanet Exploration Program.  The Green
Bank Observatory is a facility of the National Science Foundation
operated under cooperative agreement by Associated Universities, Inc.

\facility{Green Bank Telescope}\facility{Orville L. Chapman Science Learning Center at UCLA}

%
%
%

%
\bibliography{seti16}

\end{document}